# Fast cycling HTS based superconducting accelerator magnets:

## Feasibility study and readiness demonstration program driven by neutrino physics and muon collider needs.

H. Piekarz, B. Claypool, S. Hays, M. Kufer, V. Shiltsev, A. Zlobin

*Fermi National Accelerator Laboratory, Batavia, Illinois 60510, USA*

L. Rossi

*Università degli Studi di Milano, Dipartimento di Fisica*

*Istituto Nazionale di Fisica Nucleare (INFN), LASA lab, Milano 20133, IT*

Email: hpiekarz@fnal.gov

## *Executive summary*

Development of energy-efficient fast cycling accelerator magnets is critical for the next generation of proton rapid cycling synchrotrons (RCS) for neutrino research and booster accelerators of future muon colliders. We see a unique opportunity for having such magnets to be built on base of High Temperature Superconductors (HTS). Besides being superconducting at relatively high temperatures, rare-earth HTS tapes have shown very small AC losses compared to low-temperature NbTi superconductor cables.

Recent tests of the HTS-based 0.5 m long two-bore superconducting accelerator magnet have shown record high *dB/dt* ramping rates of about 300 *T/s* at 10 Hz repetition rate and 0.5 T *B*-field span. No temperature rise in 6 K cooling He was observed within the ~0.003 K error setting the upper limit on the cryogenic power loss in the magnet conductor coil to be less than 0.2 W/m [1]. Based on this result we outline a possible upgrade of this test magnet design [1, 2] to 2 T *B*-field in the 10 mm beam gap with the *dB/dt* ramping rates up to 1000 *T/s*. The power test results of this short sample magnet will be used to project both cryogenic and electrical power losses as a function of the magnet *B*-field and the *dB/dt* ramping rates. Then these projections will be scaled to the range of expected accelerator magnet beam gaps and B-fields for the proton and muon RCS accelerators.

We invite collaborators to join these studies and call for support of the R&D program aimed at comprehensive demonstration of this approach that includes design, construction, and power tests of a long prototype of the HTS-based fast-cycling accelerator magnet by 2028.

## *Table of contents:*

## *Introduction*

Next generation facilities such as muon colliders [3,4], future circular colliders [5,6] and high-intensity proton synchrotrons for neutrino research [7,8,9] demand very fast cycles of beam acceleration which in turn call for the fast-cycling accelerator magnets with the *dB/dt* of the order of tens to hundreds of *T/s*. The normal conducting magnets – for example, in the JPARC 3 GeV proton rapid cycling synchrotron (RCS) - can operate with *dB/dt* rates of up to 70 *T/s* [10] but the resistive power loss in the conductor and the magnetization loss in the magnetic cores steel make them prohibitively power inefficient for the large-scale accelerator. The need for the accelerators of high average beam power and high repetition rates have initiated studies of the fast-ramping superconducting magnets using the low-temperature superconductors (LTS). The very narrow operational temperature margin (1-2 K) of the low-temperature superconductors precluded obtaining the quench-safe operations with the rates above 4 *T/s* [11,12]. The application of the HTS superconductor to construct the fast-cycling magnet power cable allows to significantly expand operational temperature margin by operating the HTS conductor at temperature much below the critical one. This is achieved by having the total cross-section of the HTS superconductor strands sufficiently large to carry the design transport current up to 30 K. Therefore, with the operational temperature set to 5 K there is a safety margin of 25 K facilitating in this way temperature-based quench detection and protection systems. Contrary to the normal conducting accelerator magnets the superconducting cable allows to strongly minimize magnet power cable size and mass and and as a result also the size and mass of the magnetic core strongly minimizing power losses due to magnetization of the magnet core as well.

## *Test magnet design, power test arrangement and results*

The high current density of the superconductor leads to a very small cross-section of the magnet power cable and its arrangement as a narrow structure allowing to place it within the core cable space where the B-field descending from the core is minimal (Fig.1) strongly minimizing cable power losses induced by the fast-cycling magnetic field.

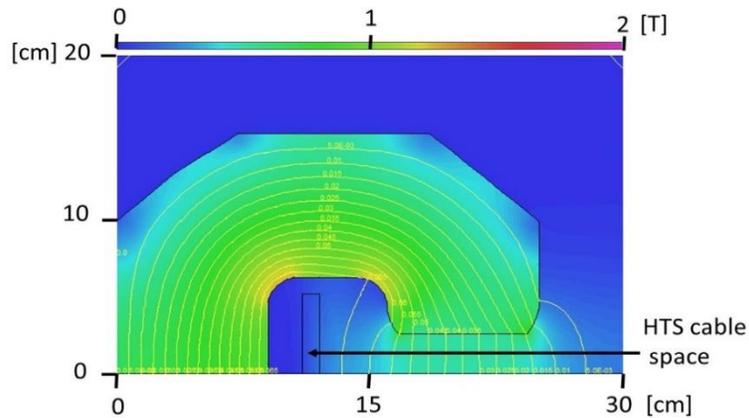

Fig. 1 B-field simulation for super-ferric magnet of 1 T in 40 mm x 100 mm beam gap.

As the beam particle decays (muons) and the synchrotron radiation (electrons) induce significant beam losses the acceleration period should be as short as possible. For the collider operation it is best when both beams are accelerated and injected to the collider ring at the same time. For these reasons we chose magnet design with two beam gaps arranged in the vertical plane and energized by the single conductor coil (Fig. 2). In such an arrangement the magnetic fields in the upper and lower gaps are of the same value but of the opposite polarities that makes it uniquely beneficial for acceleration of two beams at once using the unipolar current waveform for the beams of opposite charge particles (electrons and positrons, positive and negative muons, protons and antiprotons), or the bipolar current waveform for the same charge particles (protons). In either case the two beams will circulate in the same direction in each of the magnet gaps facilitating use of the common accelerating RF system. Also, of importance for the particle acceleration application is that in such a design, the ever-existing particle beam losses and the decay products which have lower energy than the primary beams bent out and are sent away from the HTS conductor, thus minimizing its exposure to the beam induced radiation and greatly easing requirements for the particle collimation and radiation protection systems. The 3-part magnetic core facilitates installation of the magnet conductor coil shown in Fig. 3. The 3-turn conductor coil cuts by factor 3 the required energizing current helping construction of the magnet current power supply though at the expense of the 9-fold increased inductance. The conductor coil is constructed of two sub-cables allowing to extend magnet power cable in the vertical plane while being narrow in the horizontal direction as required for minimization of the B-field crossing the cable space. For the test magnet only two 2.5 mm wide and 0.1 mm thick HTS strands [13] were helically wound at

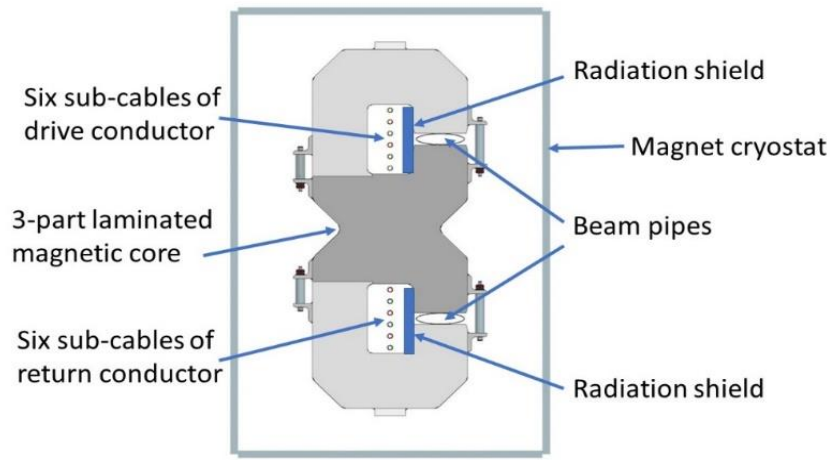

Fig. 2. Dual-bore HTS-based accelerator magnet design with warm core and cold conductor coil.

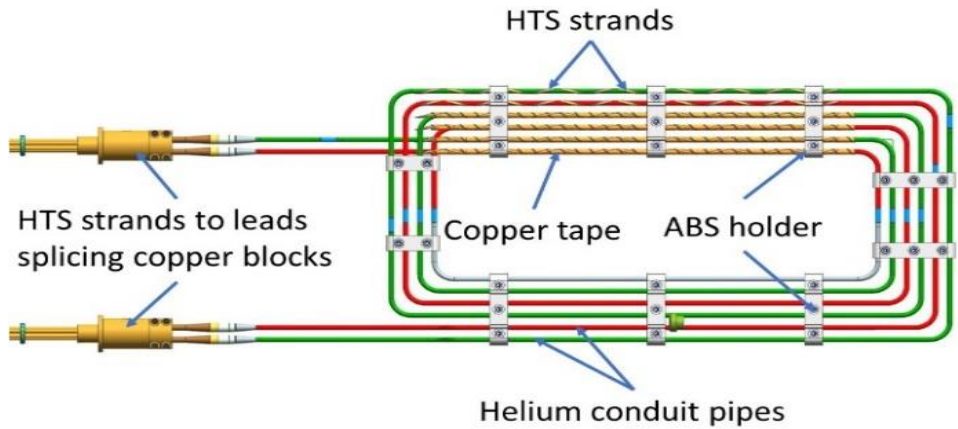

Fig. 3. HTS magnet conductor coil.

10 cm pitch on the surface of the 316LN helium conduit pipe of 8 mm ID, 0.5 mm wall. A single layer of 0.1 mm thick and 12.5 mm wide OFHC copper tape is wound helically over the strands securing their firm attachment to the helium conduit pipe. The ABS holders keep the conductor assembly together and isolating it from the magnet core. The 40-layers of MLI are wrapped over the entire cable structure between the ABS holders. The view of the test magnet assembly with the current leads is shown in Fig. 4. The conventional leads are 1.6 m long to allow for the temperature gradient from 5 K at the cold end to room temperature at connection to the power supply. Of importance, the liquid helium volume in the conductor coil within the magnet of 1 m length is 0.5 L, so a 1000 m long magnet string requires 500 L of liquid helium inventory-equivalent to the 1 standard 500 L Dewar. For convenience the current leads for test magnet are in line with magnet core but in the accelerator system they will be bent by $180^0$ at the exit of magnet cryostat and placed in the space parallel to the magnet string.

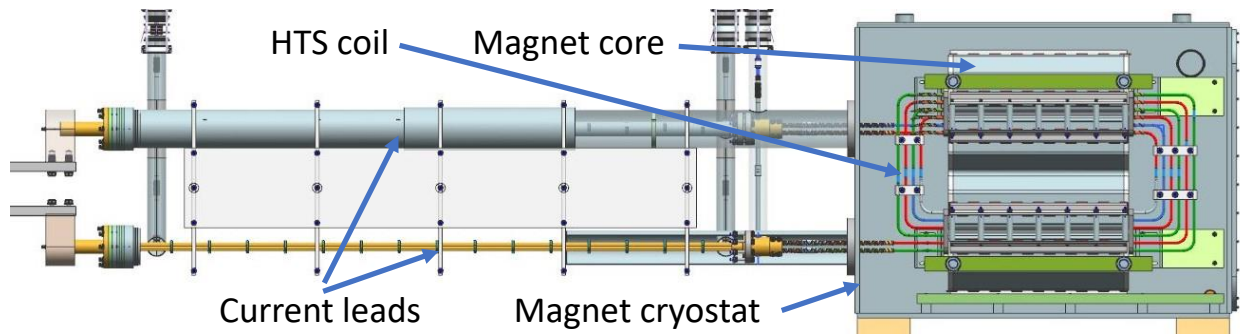

Fig. 4. Engineering design of HTS test magnet with its current leads.

The AC current source was based on the capacitor bank (20 x 960 µF) charged to 80 V and could discharge current up 1000 A. The power supply operated in the unipolar and bipolar modes. The maximum B-field was 0.377 T, and 0.5 T in the unipolar and bipolar mode of the excitation current. The full discharge current pulse length was 0.010 s with the peak magnetic field at about 0.003 s. The current pulse repetition rates were (1-10) Hz. The B-field and dB/dt responses for the unipolar and bipolar excitation currents are shown in Fig. 5 and Fig. 6, respectively. The maximum dB/dt rates are 274 T/s and 289 T/s while the *dB/dt* rates for the maximum B-field are 160 T/s and 170 T/s for the unipolar and bipolar current waveforms, respectively. The temperature change between the inlet and outlet of the magnet conductor at 1000 A excitation current operating at 1 Hz and 14

Hz repetition rates was determined to be within the +/- 0.003 K measurement error. For the liquid helium of 6 K, 0.28 MPa pressure and the flow rate of 2.4 g/s used for the magnet power tests this temperature margin corresponds to the heat loss of less than 0.1 W. We are upgrading power supply to double the discharge voltage of capacitor bank to increase the discharge current to 2 kA which will increase the B-field to 0.8 T and the dB/dt rate to about 600 T/s. With this upgrade cryogenic power losses for the test magnet will be determined with a higher precision.

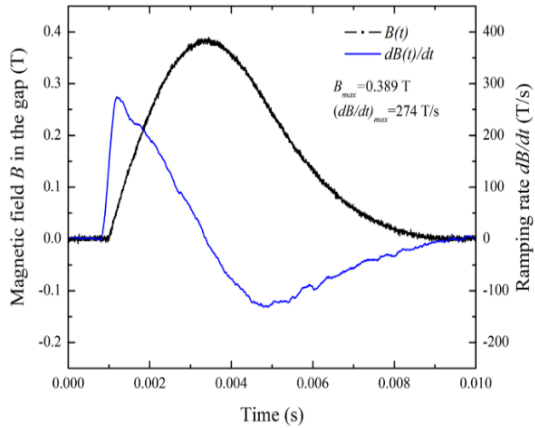 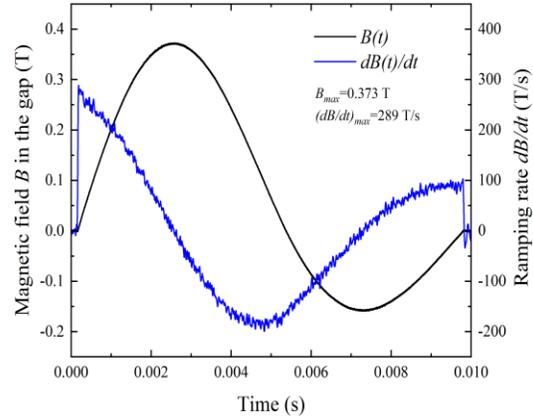

Fig. 5. *B*-field and *dB/d*t unipolar waveforms      Fig. 6. B-field and *dB/dt* bipolar waveforms

## *Test magnet upgrade to 2 T @ 500 Hz operation*

Particle acceleration requires linear response of magnet beam gap B-field to the energizing current for the equal acceleration rumps on the way to the top energy. The magnetic core saturation in the super-ferric magnet sets the limit on the usable B-field range, and magnet must be energized to the full magnetic saturation to generate the maximum range of linear response. With the Fe3%Si laminations used for the HTS test magnet the full B-field saturation is at about 2 T making B-field linear response to the energizing current up to 1.7 T. To achieve the core B-field saturation at 2 T in the test magnet with 10 mm beam gap the energizing current has to be increased 5-fold, to 10 kA-turn at 30 K. The present design of the test magnet conductor coil allows to place 5 times more HTS strands in a single layer over the cold pipe, so the test magnet core can be saturated with such an increase of magnet current. The projected beam gaps, however, for proton [9] and muon [14]

RCS accelerators are 50 mm and 25 mm, respectively. For the proton RCS the expected maximum B-field is 1 T so its energizing current is the same as for the muon RCS with 2 T B-field. The 12-fold increase of the operational current relative to the test magnet is needed for both applications. This upgrade can be achieved by using the dual layer of the HTS strands placed on the cold pipe of only slightly enlarged diameter (from 8 mm to 10 mm). The second layer of the HTS strands will be wound at the opposite angle with respect to the strands in the first layer to minimize the power losses due to self-field coupling between the strands in the opposing layers. The tentative analysis of the hysteresis and eddy losses for the HTS cable components of the test magnet indicates that to be within the 0.1 W upper limit on the heat loss the cable would have to be exposed to an average B-field of about 0.02 T, or 5% of the 0.4 T in the beam gap. With this assumption the cable component relative contributions to the heat loss are estimated as follows: Cryogenic pipes 70 %, HTS strands 19 % and Cu tape 11%. For the test magnet the dominant power loss is due to cryogenic pipes. The 0.5 mm wall of 9 mm OD SS pipe used in the test magnet can be safely reduced to 0.1 mm with the allowable working pressure of 30 bar, exceeding 10 times the operational helium pressure of 3 bar. The working pressure of 30 bar creates sufficient safety margin for the operation of the quench detection and cable protection systems. The relative reduction of the cryogenic power loss due to cold pipes will partially compensate for the power loss rise due to the 12-fold increase in the number of strands. The test magnet power supply system is based the discharge of the capacitor bank (20 x 960 µF) charged to 80 V. The discharge current of 1000 A energizes magnet to 0.4 T B-field. The increase of the energizing current to 12000 A can be achieved by the combination of: (1) - Increased discharge voltage to 480 V (magnet current coil can be designed to safely withstand more than 1000 V), and (2) - Doubling the capacitor bank to (40 x 960) µF. Also, the IGBT semiconductor technology used in the fast-switching power supplies will be considered as possible application for the HTS-based fast-cycling magnet power supply. The test magnet uses conventional (copper) current leads to connect magnet conductor coil at 5 K to power supply ends at room temperature. A tandem of the superconducting and normal conducting leads will be considered to reduce the leads power loss.

## *Long-length magnet design/construction/tests and engineering optimization*

The long-length HTS fast-cycling magnet design will be exactly as that of the test magnet with exception of the increased beam gap height and a larger number of strands. The 3-part magnet core strongly facilitates assembly of the conductor coil within the core space. The assembly of the conductor coil with two layers of HTS strands wound over the supporting cryogenic pipes can be done manually for the few long-length test magnets. For the large-scale magnet fabrication, however, the wounding of the HTS tapes over the supporting cryogenic pipes should be automated with well controlled HTS tape tension to prevent their breaking. The broken tapes can be reconnected by splicing the overlapping short strip of the HTS tape but such a process will slowdown magnet production. The long length magnet cores would be of the combined-function type to minimize the number of quadrupoles and correcting magnets in the accelerator but the skewed magnet poles will not affect conductor coil installation within the core space. The long prototype magnets will be tested for the cryogenic and electrical power losses as a function of the magnetic field and the *B*-field ramping rates. In summary, there are no major obstacles for the construction of long magnets and their large-scale fabrication. An outline of a tentative timeline for the fast-cycling HTS-based long accelerator magnet design, tests and engineering optimization is given in Table 1 below.

Table 1.

| *Item* | *Time [y]* |
|---|---|
| Short-sample accelerator magnet | 2 |
| Long accelerator magnet prototype | 4 |
| Engineering optimization | 2 |